# Title: Hydrogen-induced lattice cohesion weakening favors atomic displacement

**Authors:** L. Gao[1,2]*†, Y. Mao[1], M. Wilde[3], X. Yi[4]*, C. Li[1,5], S. Wang[6], T. Schwarz-Selinger[7], J. Coenen[1,2], R. Kembleton[2], S. Brezinsek[1], Ch. Linsmeier[1] and G.-H. Lu[8]*

**Affiliations:**

[1]Institute of Fusion energy and Nuclear waste management − Plasma physics (IFN-1), Forschungszentrum Jülich GmbH; Jülich, 52425, Germany.

[2]Gauss Fusion GmbH; Garching, 85748, Germany.

[3]Institute of Industrial Science, University of Tokyo; Komaba 4-6-1, Meguro-ku, Tokyo, 153-8505, Japan.

[4]School of Materials Science and Engineering, University of Science and Technology Beijing; Beijing, 100083, China.

[5]Shanghai Institute of Applied Physics, Chinese Academy of Science; Shanghai, 201800, China.

[6]State Key Laboratory of Nuclear Physics and Technology, School of Physics, Peking University; Beijing, 100871, China.

[7]Max-Planck-Institut für Plasmaphysik; Garching, 85748, Germany.

[8]School of Physics, Beihang University; Beijing,100191, China.

*Corresponding authors: li.gao@extern.fz-juelich.de (L. Gao), xiaoouyi@ustb.edu.cn (X. Yi), lgh@buaa.edu.cn (G.-H. Lu)

†Present address: Gauss Fusion GmbH, Parkring 29, 85748-Garching, Germany

**Abstract:** Atomic displacement – the fundamental process underlying diverse deformation and damage phenomena in metals, from irradiation defect production to stress-driven dislocation motion − is governed by interatomic cohesion strength. Here, lattice-dissolved hydrogen (LDH) occurring in metals under direct hydrogen exposure is identified to effectively weaken lattice cohesion, and thereby facilitating atomic displacement and dislocation movement upon plastic deformation in sub-threshold stress regime. This atomic-scale insight provides a physically transparent mechanism for hydrogen-enhanced localized plasticity implicated in hydrogen embrittlement. We quantitatively verify the hydrogen-induced lattice cohesion weakening effect on metal surfaces exposed to low-energy hydrogen plasma, where massive defects are generated despite the absence of sufficient ion momentum for direct displacement damage. By unprecedentedly quantifying the cohesion-weakening effect of LDH independently from defect-trapped H, we establish a new paradigm to understand hydrogen embrittlement.

**Main Text:**

Hydrogen (H), the most promising carbon-neutral energy carrier, is a ubiquitous impurity in crystalline solids due to its small atomic size and high chemical reactivity. Interacting with either crystal defects or lattice atoms, and referred to as defect-trapped or lattice-dissolved H (LDH[1]) respectively, hydrogen profoundly modifies the electronic and mechanical properties of host metals (*1*). One well-known adverse effect of hydrogen incorporation is H embrittlement (HE), which strongly degrades the mechanical stability of metals, especially high strength/stiffness body-centered-cubic (bcc) metals and alloys with a wide range of engineering applications. A serious concern in industry and hot topic in materials research, HE ranks among the key challenges in developing safe and cost-effective hydrogen storage and transportation systems for a carbon-neutral 'hydrogen economy'. Elucidating HE mechanism is highly important, yet despite a 150-year-long quest the fundamental physics of HE still remain elusive. As early as 1875, Sir Johnson (*2*) conjectured that diffusive H (i.e., LDH) causes metal embrittlement, but this hypothesis is challenging to validate as LDH in metals is extremely difficult to detect. Consequentially, HE research has turned to focusing on defect-trapped H being easier to assess experimentally; almost all proposed HE models (*3*-*6*) do not differentiate defect-trapped hydrogen from LDH but attribute ambiguously to 'dissolved H' in metals. The well-documented hydrogen-enhanced localized plasticity (HELP (*6, 7*)) effect is one such example, where H 'dissolution' is considered to affect dislocation motion and localized plasticity in metals. In this article, we argue there is rarely any stably-trapped hydrogen at dislocations in samples tested under HELP-relevant conditions, hence LDH is the primary player in enhancing dislocation mobility in metals undergoing plastic deformation. The energy barrier of dislocation movements is determined by Peierls potential that derives directly from the periodic lattice structure and depends sensitively on the interatomic bond/cohesion strength (*8*). To quantitatively evaluate the LDH effect on the interatomic cohesion strength, we introduce herein another surrogate experimental setting, where strong lattice distortion is produced in a metal surface exposed to a low-energy H plasma. Impacting H ions create massive vacancy-type defects in the immediate surface vicinity upon LDH population with otherwise insufficient ion energy, in analogy to imposing sub-threshold stress that causes dislocation motion in LDH-populated samples under HELP conditions. We postulate that in both scenarios, the diffusive LDH-induced lattice cohesion weakening promotes an enhancement of atomic displacements in the host metal.

[1] In this work, the term 'LDH' stands for all mobile H atoms occupying/jumping among interstitial sites in the metal matrix, regardless of the kinetic energy they are carrying.

**Lattice dissolved hydrogen-enhanced localized plasticity**

The HELP mechanism was proposed by Beachem (*9*) in 1972 and realized through *in situ* transmission electron microscopy (TEM) experiments by Birnbaum, Robertson and coworkers *(6, 7),* which enabled the first direct observation of H-localized plastic deformation. The introduction of $H_2$ into a TEM equipped with an environmental cell (E-cell) caused enhanced mobility of dislocations in almost every metal under externally applied sub-threshold stress (*10*). With increasing $H_2$ pressure the dislocation velocity increased rapidly, while removing $H_2$ gas from the E-cell terminated the dislocation motion (*11*). More recently, the HELP mechanism was demonstrated to apply also to the bowing-out of a single screw dislocation in α-Fe (*12*). Importantly, H-enhanced dislocation mobility or H softening has been observed almost exclusively by *in situ* TEM, in stark contrast to the conventional metallurgical paradigm of solute-induced hardening (dislocation pinning (*8, 14*)).

To rationalize HELP, one finite-element study proposed that H shields elastic interactions among multiple dislocations, thereby enhancing their mobility (*15*). However, this continuum-level description lacks an atomistic basis and thus cannot account for the bowing-out of an isolated dislocation (*12*). Subsequent atomistic simulations sought to reconcile both H-induced hardening and softening within unified frameworks (*16–19*), whereas simulations focusing solely on HELP cannot reproduce H-enhanced dislocation mobility (*20*). To date, only Lu *et al.* (*21*) have reproduced HELP by explicitly placing hydrogen at interstitial lattice sites in *ab initio* calculations. Experimentally, recent *in situ* studies imaging dislocations in bcc metals under cathodic charging have reported conflicting H effects, ranging from enhanced mobility (*22, 23*) to pronounced pinning (*22, 24, 25*). The decisive factor governing these disparate observations − and thus the underlying physics of H-enhanced dislocation mobility − remains unresolved.

During *in situ* TEM experiments (*6, 7, 10, 12*), H is loaded into samples under electron beam irradiation via electron-induced dissociation of gas molecules at the surfaces (Fig. 1A). For a TEM operated typically at 200 kV, the incident electrons can impart kinetic energy on the order of 100 eV to the recoiled H atoms. In contrast, de-trapping energies of H in dislocations are typically only a fraction of an eV (*26*). As a consequence, dislocations act only as temporary traps for incorporated H under electron bombardment because energetic electrons can easily de-trap H into the lattice (*26*) and make almost all H diffusive. Accordingly, LDH emerges as the primary contributor to H-enhanced dislocation mobility. This interpretation is consistent with the rapid increase in dislocation velocity observed upon elevating the $H_2$ pressure, as well as the immediate suppression of mobility following gas removal *(11)*. However, the potential role of electron-beam irradiation during *in situ* TEM − specifically its ability to mobilize otherwise weakly bound, dissolved H − has been largely overlooked. Similarly, atomistic simulations of H-dislocation interactions in α-Fe (*20, 24, 25*) have assumed stable H trapping at dislocation cores. This assumption is unlikely to hold under electron irradiation, where weakly trapped H would be readily released from the core region, potentially accounting for the inability of these studies to reproduce

HELP effects. As trapped hydrogen at dislocation cores has been experimentally demonstrated to favor dislocation pinning and crack nucleation (i.e., embrittlement) in our parallel work *(28)*, we further propose that the coexistence of LDH and defect-trapped hydrogen in the metal matrix underlies the experimentally reported dual role of hydrogen (*22, 23*) in either enhancing dislocation mobility or inducing pinning. We are motivated hereafter to describe the relevant physical mechanisms and further quantify them based on H supersaturation close to free surfaces after low-energy plasma exposure.

**H-induced lattice cohesion weakening (HiLaCoW) effect**

An LDH atom imposes lattice distortion via anisotropic volume expansion (*29*) and accordingly weakens the local interatomic cohesive strength, applying to both perfect bulk crystal and the intrinsically under-coordinated structure surrounding crystal defects. Taking a screw dislocation in bcc metals as an example, Fig. 1 schematically illustrates the decisive role of electron-mobilized LDH atoms in enhancing dislocation mobility upon low applied stress during HELP experiments. Upon electron impact, H atoms de-trap from the dislocation core with a sizeable amount of kinetic energy and will thus appear transiently at interstitial sites between the core region and the matrix (hexagonal ring in rose color in Fig. 1B). LDH-weakened interatomic cohesion favors atomic displacement upon applied stress. In the low stress regime where dislocation motion is governed by kink-pair nucleation, the LDH-facilitated atomic displacement accounts for the lower kink-pair formation enthalpy (*17*) and thus a higher nucleation rate per unit length of the dislocation (Fig. 1A regime III). Effectively, the velocity of a moving dislocation line is proportional to the number of nucleated kink pairs (see the equation in Fig. 1A) and is hence enhanced due to the transient presence of LDH. We further interpret the H-enhanced bowing-out of a single dislocation (*12, 30*) by a sequential formation of many kink pairs on one single primary dislocation line in the presence of multiple LDH atoms. Critically, H-enhanced dislocation mobility (or HELP) arises *not* from a universal reduction in lattice resistance but from a selective weakening of lattice cohesion that lowers the activation barrier for kink-pair nucleation. Consequently, LDH promotes dislocation motion only in sub-threshold stress regimes where glide requires additional thermal activation, as realized by *in situ* applying external stress to specimens immersed either in $H_2$ atmosphere upon TEM examination *(6, 7, 12)* or under cathodic charging (*23*).

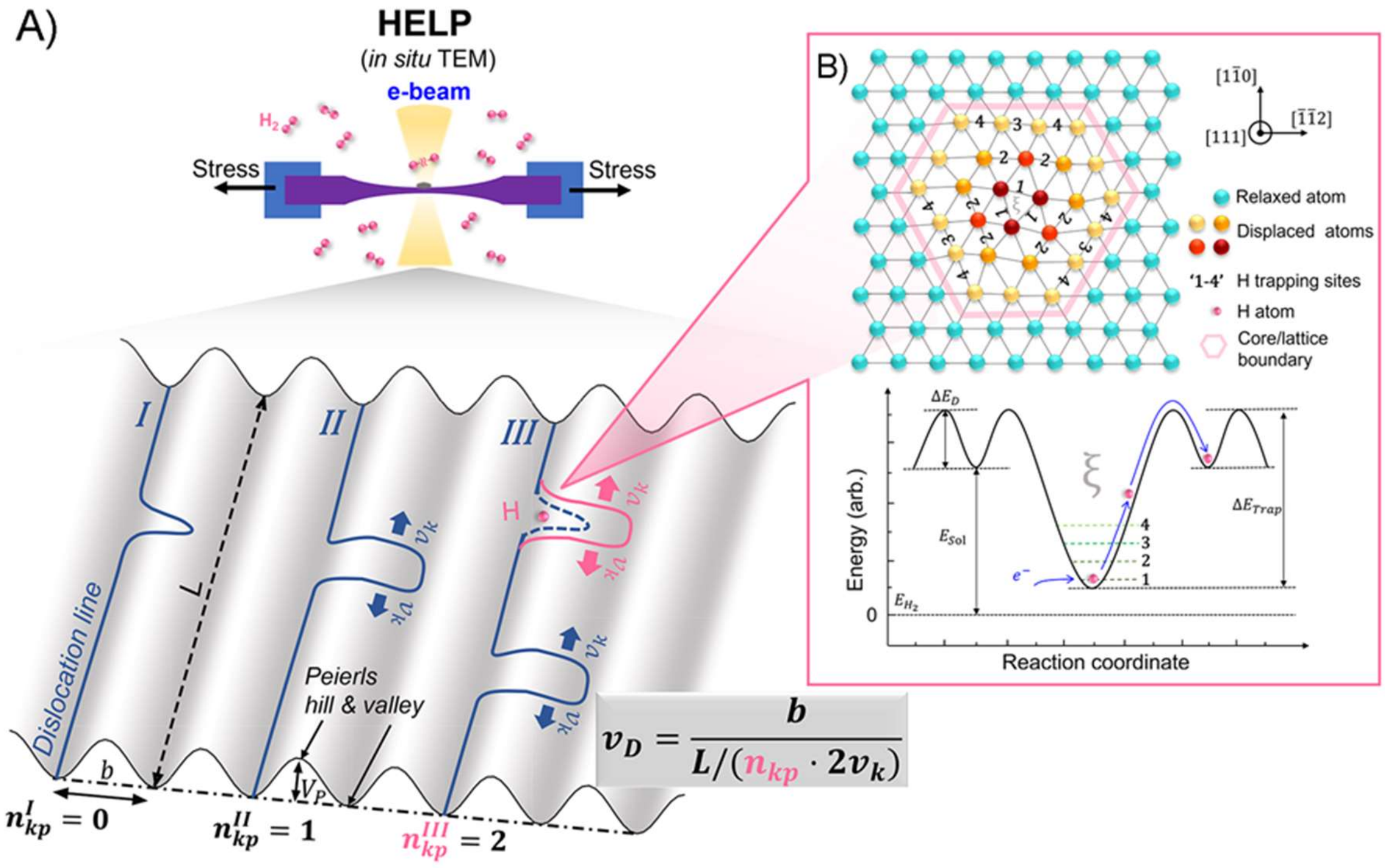

***Fig. 1. Schematic illustration of enhanced kink-pair nucleation in low stress regime due to the transient presence of interstitial H between the dislocation core and matrix under TEM electron beam irradiation.*** *A) Taking a screw dislocation in a bcc metal as an example, three different regimes of dislocation motion are illustrated on different dislocation lines. I) The dislocation resolved shear stress (RSS) from the applied load is too low to initiate any well-separated kink pair formation, and the dislocation line remains stationary; II) The RSS is higher than the*

*threshold for the generation of well-separated kink pairs. III) The transient presence of interstitial H atoms, as mobilized by energetic electrons bombarding H-decorated dislocations, allows the nucleation of well-separated kink-pairs at initially unavailable sites and these H atoms do not drag the newly-generated kink pairs via trapping. This enhances the kink-pair nucleation rate and in turn the dislocation mobility, which is proportional to the number of well-separated kink pairs ($n_{kp}$). B) Top: the core structure of a screw dislocation viewed from **z** = [111] with possible H trapping sites indicated (see Fig. 6 in Ref. (24)); Bottom: energy potential diagram of H at a screw dislocation core in the bulk interior of a metal, where the energy for a $H_2$ molecule ($E_{H_2}$), H in solution ($E_{sol}$) and the activation energy for H diffusion ($\Delta E_D$) and de-trapping ($\Delta E_{trap}$) from the screw dislocation ($\xi$) are indicated. The inset equation defines the dislocation velocity ($v_D$).*

The LDH-favored atomic displacements under external stimulation is the essential feature common to the enhanced dislocation motion under HELP conditions and to the defect production in metal surfaces under low-energy H plasma/ion irradiation (see, e.g., the vacancy-type nanocavities in Fig. 2). Both scenarios involve a strikingly enhanced lattice atom mobility caused by the presence of LDH at the adjacent interstitial site: LDH facilitates atomic displacement and kink-pair formation under HELP conditions with subthreshold applied stress; while during low-energy H exposure lattice distortion can be generated by impacting H ions with otherwise insufficient energy via a synergistic process with LDH atoms in the immediate surface vicinity (*31*). We will term in the following the proposed mechanism as hydrogen-induced lattice cohesion weakening (*HiLaCoW*) effect. For bcc tungsten (W), the exposure to low-energy H plasma causes the formation of hydrogen supersaturated surface layers (HSSL) composed of astonishingly large numbers (~10 at. %) of hydrogen captured by irradiation-induced vacancy-type defects (*31-33*). Interestingly, the H-occupied nanocavities in the early-stage HSSLs (*31*) evolve substantially upon electron beam irradiation (see also Fig. 2A→B), similar to HELP experiments where significant redistribution of incorporated H occurs during TEM examination. Remarkably, the collisional energy transfer threshold for HSSL formation amounts to only 5 eV (*32*) – considerably smaller even than the W lattice cohesion of 8.9 eV/atom *(34)* – which we consider as a further experimentally well-established manifestation of the proposed HiLaCoW effect, in addition to HELP. Therefore, measuring the HiLaCoW effect upon HSSL formation on low-energy plasma-exposed W surfaces offers atomistic understanding of HELP effects and also an unprecedentedly quantifiable assessment on the role of LDH in metals.

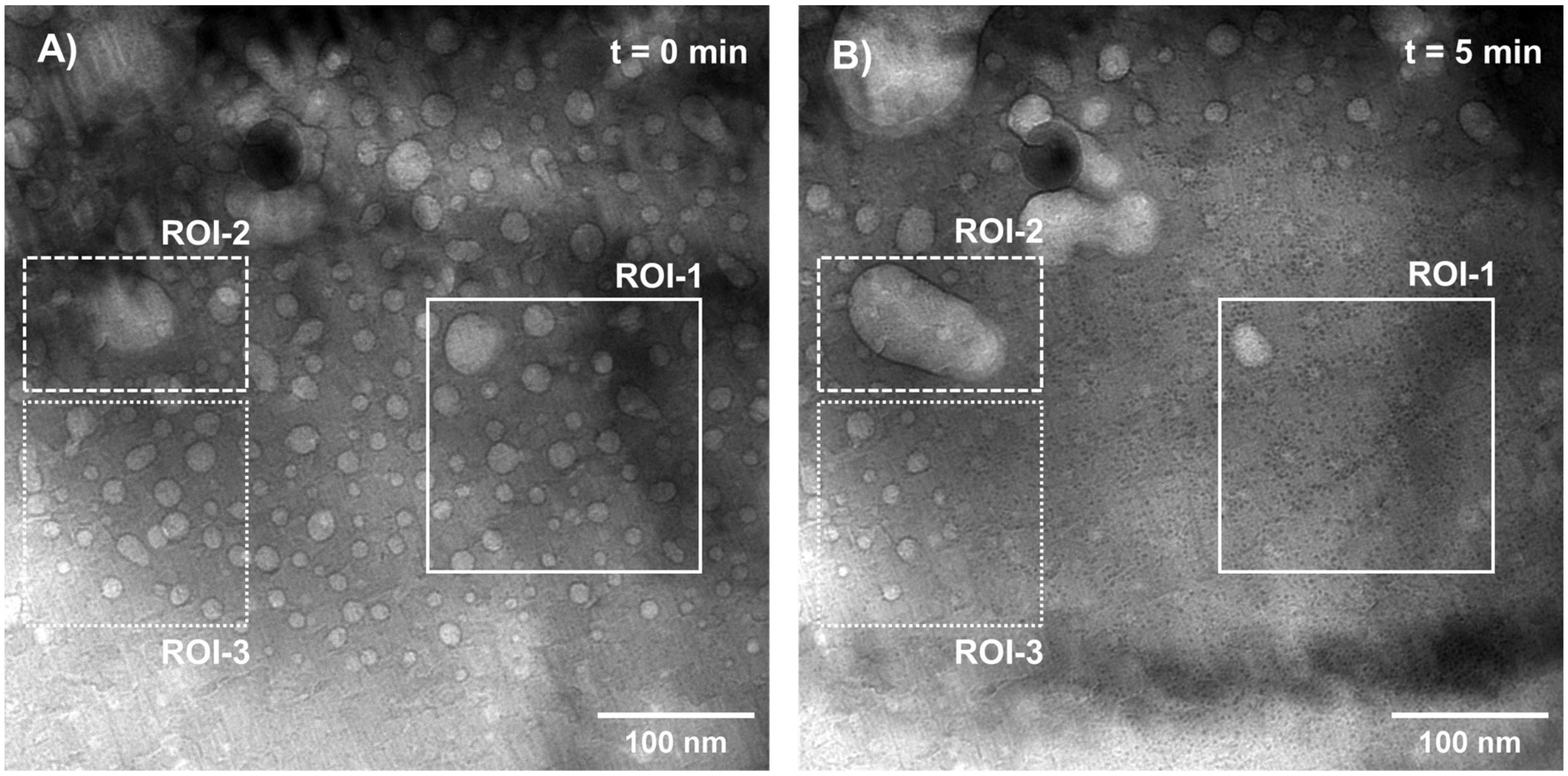

***Fig. 2. In situ TEM observation of vacancy-type nanocavities on a H plasma-exposed tungsten surface demonstrates a strong e-beam effect on the H distribution within the H supersaturated surface layer (HSSL).*** *Micrographs were captured at room temperature under kinematical bright-field condition, at A) t = 0 min and B) t = 5 min of e-beam irradiation, using an image exposure time of less than 0.5 s. ROI-1 highlights the shrinkage and disappearance of nanocavities, ROI-2 exemplifies a clear growth in cavity size via coalescence; while ROI-3 shows the heterogeneous evolution of nanocavities under the electron beam.*

**Quantifying HiLaCoW effect upon HSSL formation**

In metals under irradiation, lattice atoms are displaced in collisions with incident particles creating point defects of self-interstitial atoms (SIA) and vacancies, known as Frenkel pairs. In a perfect lattice, a stable Frenkel pair (s-FP) forms only if the SIA is driven into an interstitial site with distance beyond the spontaneous FP recombination radius *(35)* to the correlated vacancy. The s-FP creation involves a high energy barrier ( $E_{d_0}^{s-FP}$ , see Case II, Fig. 3) because of the energy-intensive insertion/squeezing of SIA into the compact matrix. For tungsten, $E_{d_0}^{s-FP}$ amounts to 40-70 eV, depending on the crystallographic orientation *(35,36)*. Collisions transferring less kinetic energy create transient Frenkel pairs (t-FPs) of shorter separations that are intrinsically unstable, i.e., susceptible to SIA-vacancy recombination. However, if an effective SIA absorption sink exists nearby the collision site (such as, the free surface or extended defects like grain boundaries or vacancy clusters (*37*)), even t-FPs can materialize as surviving vacancies. Only a short-distance SIA translation to the absorption sink rather than the energy-intensive SIA squeezing into perfect lattice is involved. Such materialized vacancies consume considerably less energy than in the perfect bulk crystal (Case III, Fig. 3). This accounts for the significantly reduced energy threshold of vacancy-type defects production upon HSSL formation (*31-33*), where the plasma-exposed W surfaces act as effective SIA sink for vacancy materialization. We denote $E_{d_0}^{t-FP}$ as the significantly decreased energy threshold of t-FP creation upon an effective SIA sink, with respect to $E_{d_0}^{s-FP}$ (see Fig. 3C). Notably, in the vicinity of the free surface acting as SIA sink (and possibly near extended defects in the bulk), the inherently reduced geometric coordination may lead also to enhanced flexibility of lattice atoms towards atomic displacements, as manifested by the spontaneous low-temperature reconstruction of the W(100) single crystal surface (*38*). However, such inherent subsurface reconstruction does not apply to all the W grain orientations (*38*), so we neglect herein the influence of such undercoordination in the defect vicinity on $E_{d_0}^{t-FP}$. In the presence of LDH in the matrix (Fig. 3B), HiLaCoW effect comes into play further reducing $E_{d_0}^{t-FP}$ to $E_{d_LDH}^{t-FP}$ towards vacancy materialization (Case I→III, Fig. 3).

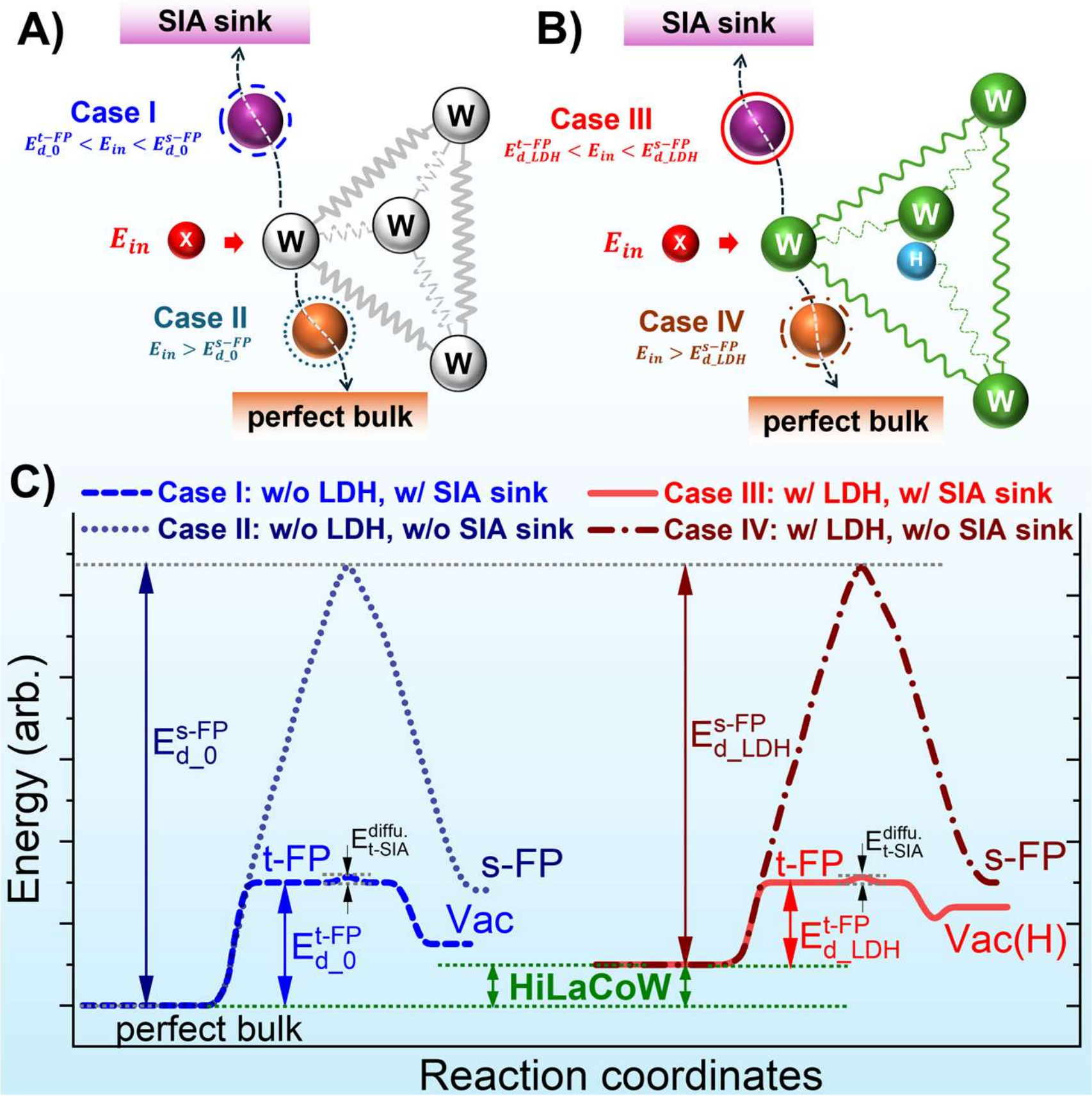

***Fig. 3. H-induced lattice cohesion weakening facilitates atomic displacement and point defect production upon external energy input ($E_{in}$).*** *In a perfect lattice (Case II), the formation of stable Frenkel pairs (s-FP) (i.e., self-interstitial atoms (SIAs) and stable vacancies) requires a large threshold in energy transfer (*$E_{d_0}^{s-FP}$*) via kinetic collisions. Collisions with energy transfer below* $E_{d_0}^{s-FP}$ *create transient Frenkel pairs (t-FPs) of shorter separations that are still susceptible to SIA-vacancy recombination. However, if the t-FPs exist in close vicinity of pre-existing defects (i.e., the free surface) that can act as SIAs sink effectively absorbing SIAs, the vacancies from t-FPs can survive after the highly mobile SIAs transfer to the defect sink (Case I&III). The high*

*mobility of SIAs is indicated by the extremely low activation energy of SIA diffusion ($E_{diff}^{SIA}$) in C). In the presence of LDH (Case III&IV), the lower limit of t-FPs formation threshold ($E_{d_LDH}^{t-FP}$) is further reduced due to the H induced lattice cohesion weakening (HiLaCoW) effect, as experimentally demonstrated in Ref. (31).*

In principle, HiLaCoW should influence both $E_d^{s-FP}$ and $E_d^{t-FP}$ in an LDH-occupied W lattice (Fig. 3). However, the inherent challenge on detecting such atomistic events in the bulk and the rather high value of $E_{d_0}^{s-FP}$ make it rather tricky to experimentally verify the HiLaCoW magnitude. We therefore focus herein on $E_{d_LDH}^{t-FP}$ upon early-stage HSSL formation (*31*) in plasma-exposed W surfaces, where materialized vacancies after H decoration (Case III, Fig. 3B&C) can be experimentally characterized. Tungsten samples were exposed to low-fluence H plasma for HSSL formation at 300-600 K (see ***Materials and Methods in Supplementary***) within the intrinsic surface action range in absorbing SIAs (*39*). HSSL propagation into larger depth upon high fluence (see ***Fig. S1 in Supplementary Section 1***) or via vacancy diffusion at T > 600 K (*40*) is avoided. By measuring H depth profiles and the thickness of HSSL formed at different temperatures, we attempted to quantitatively evaluate the HiLaCoW effect and, if any, its temperature dependence. However, H occupancy at decorated vacancies depends also sensitively on temperature. To compensate for any thermal loss of H occupancy (see ***Fig. S2 in Supplementary Section 2***) upon HSSL formation at elevated temperatures, all the HSSL samples were treated with a second H plasma with 0 V bias and a much higher H fluence at 300 K. All existing vacancies are (re)filled with H to the same occupancy level (i.e., 5 H atoms per vacancy at 300 K (*41*)).

Then, $^{1}H$-$^{15}N$ nuclear reaction analysis (NRA, see ***Materials and Methods in Supplementary***) was applied to measure H depth profiles with a nanometer resolution *(42)*, as shown in Fig. 4A.

Pristine W sample (dashed line) without HSSL formation still shows a Gaussian surface peak corresponding to absorbents, while all other samples exhibit a plateau after the surface peak indicating HSSL formation. The HSSL thickness (*32*) is estimated from the inflection point in the sigmoid fit curves to the declining edge of the H depth profiles. As plotted Fig. 4B, the HSSL thickness on polycrystalline W surfaces exhibits a monotonic decline with increasing temperature, seemingly indicating a temperature-dependent HiLaCoW effect. However, during the 2-stage vacancy materialization process HiLaCoW effect is only governing the very first stage of t-FP generation but cannot affect the subsequent SIA absorption. The action range of free surface in absorbing SIA determining the HSSL thickness is solely governed by the surface energy (*39*). We therefore adopt in Fig. 4B the surface energy profiles of the W(110), (100) and (111) crystallographic facets (*43*). The temperature-dependent evolution of the HSSL thickness reflects an ensemble average of these facet-specific surface energy reductions at 200-800 K. While the SSL thickness decreases linearly between 400 K and 600 K, the anomalous departure at 300 K coincides with the transition below the Debye temperature of tungsten ($\omega_D \approx 400$ K), where lattice phonon contributions to the surface energy significantly diminish. Such a direct thermodynamic coupling suggests that the observed temperature dependence of the HSSL thickness can be attributed to the thermally induced reduction in surface energy.

Most importantly, for W with an extremely small thermal expansion coefficient ($10^{-5}$ $K^{-1}$ (*44*)), the thermal expansion of W lattice at the applied temperatures is negligible ($10^{-3}$ from 300 to 600 K) compared with the localized LDH-induced lattice dilatation (i.e., 2.9 $Å^3$ (*29, 45*), accounting for ~10 % of the W unit lattice volume. We therefore further affirm that HiLaCoW effect is temperature-independent for the applied temperature range. The threshold energy transfer on t-FP generation in the immediate vicinity of the free surface towards HSSL formation in W was experimentally determined to be 5 ± 1 eV at 300 K (*33*). For the sake of simplicity, we herein assign the SIA formation energy (i.e., 9.6 eV *versus* 3.6 eV for a vacancy formation (*46*)) in tungsten for $E_{d_0}^{t-FP}$ as the lower threshold of kinetic t-FP formation energy. This gives rise to at least 4.6 ± 1 eV as the quantified HiLaCoW effect at the applied experimental conditions, accounting for almost half of the t-FP formation energy. Notably, this is the very first time in experimentally evaluating the cohesion-weakening effect of a diffusive LDH atom in a metal.

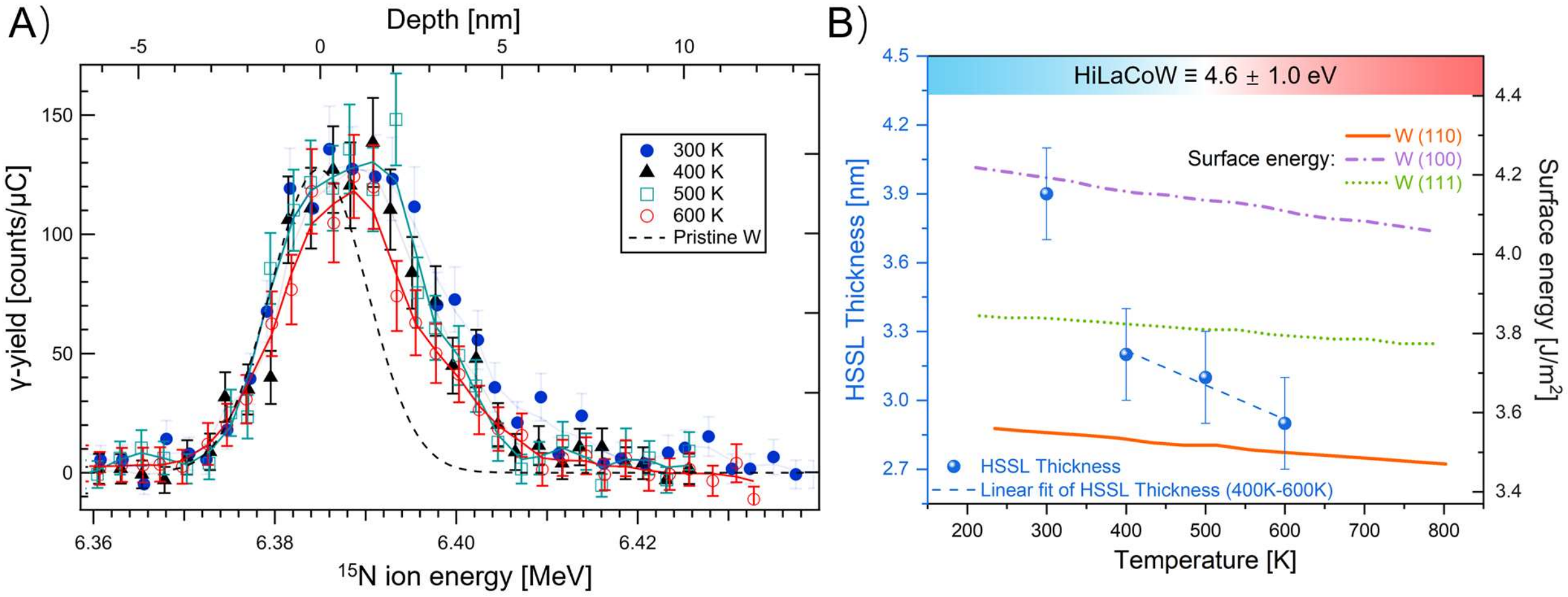

***Fig. 4. Temperature-dependent tungsten surface energy rather than temperature-independent HiLaCoW effects determines the thickness of early-stage hydrogen supersaturated surface layers (HSSL).*** *A) H depth profiles of HSSL initially formed at 300-600 K in plasma-exposed W surfaces and refilled to the same H occupancy level at materialized vacancies by a second plasma exposure at 300 K, as measured by* $^{1}$*H-*$^{15}$*N nuclear reaction analysis (NRA). B) The evolution of initial HSSL thickness due to the temperature-dependent surface energy and accordingly the action range of the free surface in absorbing SIAs; Plotted also are surface energy profiles of the W(110), (100) and (111) crystallographic facets adopted from Ref. (43). The temperature-dependent SSL thickness measured on polycrystalline W surfaces exhibits a definitive correlation with the thermal scaling of the constituent facet surface energies. The pronounced deviation of the 300 K measurement from the higher-temperature (400-600 K) linear regime reflects a fundamental shift in thermodynamic behavior below the Debye temperature of tungsten (*$\omega_D \approx 400$ *K), demonstrating that the thermal evolution of the SSL thickness is fully governed by corresponding variations in surface energy The hydrogen-induced lattice cohesion weakening (HiLaCoW) effect remains temperature-independent, amounting to 4.6 ± 1 eV as determined at 300 K in Ref.(33) and being constant for the applied conditions from 300 K to 600 K.*

It should be noted that, the LDH effect does not constitute a prerequisite for either the enhanced dislocation mobility upon HELP conditions or vacancy materialization during HSSL formation, but acts as a stimulus-dependent facilitator by weakening local lattice cohesion and lowering the energetic cost of atomic displacements. For HELP effect, this corresponds to a lower activation barrier for the thermally activated kink pair formation, thereby accelerating dislocation motion under low resolved shear stresses. At sufficiently high stresses, where dislocations overcome the Peierls barrier and glide as a whole line, dislocation motion becomes insensitive to LDH. For HSSL formation upon low incident ion energy, the HiLaCoW-favored atomic displacement is manifested as vacancy materialization after SIA absorption at the free surface. Similarly, when the incident ion energy is high enough to create directly s-FPs, the presence of LDH and the associated HiLaCoW effect becomes unnecessary. Acting as a stimulus-dependent facilitator to atomic displacements, LDH atoms in the two paralleled phenomena are sourced differently. During HELP experiments, LDH carrying sizable kinetic energy is generated from temporarily-trapped H atoms at dislocation cores by impacting electrons; while upon HSSL formation LDH with almost zero kinetic energy stems from thermalized ions after implantation. Nevertheless, how LDH is produced during the process is irrelevant to the here-quantified HiLaCoW effect as the common essential of atomic displacements in the two seemingly unrelated phenomena.

To summarize, we propose that HELP effects in metals are governed by the transient presence of LDH surrounding dislocations, being generated either indirectly by the electron beam irradiation in TEM with $H_2$ atmosphere or directly by *in situ* H loading (e.g., cathodic charging (*23*) or plasma exposure (*31-33*)). Via promoting a higher nucleation rate of kink pairs without slowing down the formed kinks through trapping, LDH effectively enhances dislocation mobility upon plastic deformation in low or medium stress regime. Such LDH-favored atomic displacements upon external stimulation via the here-defined HiLaCoW effect is also identified as the common essential of sub-threshold defects production upon HSSL formation to the enhanced dislocation mobility under HELP conditions. By paralleling the two seemingly irrelevant scenarios, we achieve for the first time a quantitative evaluation of the HiLaCoW effect. The HELP model is suggested to be applicable for interpreting results of *in situ* HE testing experiments where direct H exposure can ensure a sufficiently high LDH population in the specimens. Otherwise, LDH concentration is negligibly low, making the direct observation of H-enhanced dislocation mobility in metals exclusive to *in situ* TEM experiments rather than commonplace. Instead of attributing HE ambiguously to the diffusive LDH as in conventional theory, the present work uncovers LDH in governing the well-documented HELP mechanism and offers unprecedented quantification of HiLaCoW effect. In parallel, dislocation pinning and embrittlement in non-hydride-forming metals was demonstrated to be induced by defect-trapped H at dislocations (*28*). Hence, LDH and defect-trapped H are playing their unique, distinguishable roles on impacting dislocation motion in metals upon external stress application. The conversion from defect-trapped H to LDH may be able to provide a simple but effective mitigation of dislocation-oriented embrittlement of metals subjected to H environments. Identifying the unique role of LDH from that of generally dissolved H in enhancing dislocation mobility via lattice cohesion weakening, the present work defines a novel research paradigm to understand (and ideally mitigate) hydrogen embrittlement.

**Acknowledgments:** Fruitful discussions with Dr. U. von Toussaint, Dr. Y. Ferro and Dr. D. Mason are highly appreciated. AI is employed for language polishing of some parts of the manuscript.

**Funding:**

This work receives the support from the BMFTR Fusion 2040 project SyrVBreTT (Grant No.:13F1011G).

**Author contributions:**

Conceptualization: LGO

Methodology: LGO, YMO, SWG

Investigation: LGO, MWE, XYI, TSS, CLI

Visualization: LGO, CLI, SWG, MWE, XYI

Funding acquisition: RKN, SBK, CLR

Project administration: JCN, RKN, SBK, CLR

Supervision: JCN, CLR

Writing – original draft: LGO, XYI, MWE

Writing – review & editing: all

**Competing interests:** Authors declare that they have no competing interests.

**Data, code, and materials availability:** All data are available in the main text or the supplementary materials.

# Supplementary Materials for

## Hydrogen-induced lattice cohesion weakening favors atomic displacement

L. Gao[1,2]*†, Y. Mao[1], M. Wilde[3], X. Yi[4]*, C. Li[1,5], S. Wang[6], T. Schwarz-Selinger[7], J. Coenen[1,2], R. Kembleton[2], S. Brezinsek[1], Ch. Linsmeier[1] and G.-H. Lu[8]*

**Affiliations:**

[1]Forschungszentrum Jülich GmbH, Institute of Fusion energy and Nuclear waste managemen – Plasma physics (IFN-1), 52425, Jülich, Germany;
[2]Gauss Fusion GmbH, 85748, Garching, Germany;
[3]University of Tokyo, Institute of Industrial Science, Komaba 4-6-1, Meguro-ku, 153-8505 Tokyo, Japan;
[4]University of Science and Technology, School of Materials Science & Engineering, 100083 Beijing, China;
[5]Shanghai Institute of Applied Physics, Chinese Academy of Science, 201800, Shanghai, China;
[6]State Key Laboratory of Nuclear Physics and Technology, School of Physics, Peking University, Beijing 100871, China;
[7]Max-Planck-Institut für Plasmaphysik, 85748, Garching, Germany;
[8]School of Physics, Beihang University; 100191, Beijing, China.

*Corresponding authors: li.gao@extern.fz-juelich.de (L. Gao), xiaoouyi@ustb.edu.cn (X. Yi), lgh@buaa.edu.cn (G.-H. Lu)

†Present address: Gauss Fusion GmbH, Parkring 29, 85748-Garching, Germany

**The Supplementary Materials include:**

Materials and Methods
Supplementary Text
Figs. S1 to S2
References (47-50)

## Materials and Methods

Polycrystalline, hot-rolled tungsten (W) samples with 99.97 wt.% purity (Plansee SE Austria) and dimension of 12×15×0.8 $mm^3$ were mechanically polished and subsequently electro-polished in a 1.5 wt.% NaOH solution, achieving a final thickness of ~0.7 mm and mirror surface finish. Annealing in vacuum at 1200 K for 2 hours were performed for all samples to relieve the stress induced during mechanical polishing process and to degas possible intrinsic H from the production process of the samples. The samples were then exposed to H plasma in PlaQ (*47*), a low-temperature electron-cyclotron-resonance plasma facility housed at the Max-Planck-Institute for Plasma Physics in Garching, Germany. One set of 4 as-treated W samples were plasma-exposed to low H fluence of $5\times10^{20}$ $H^+\cdot m^{-2}$ (*47*) at 300, 400, 500 and 600 K, respectively. The plasma ion flux was measured to be ~$1.1 \times 10^{20}$ H $m^{-2}$ $s^{-1}$. The ion flux from the plasma consisted of ion species $H_3^+$ (94 %), $H_2^+$ (3 %) and $H^+$ (3 %), each contributing 3 H, 2 H and 1 H atom to the total H fluence, respectively. Together with the -15 eV plasma potential, the applied -400 V bias voltage result in an ion energy of 415 eV. Only for $H^+$ ions the energy per H corresponds to the full ion energy (415 eV) upon impact, hence only $H^+$ ions are supposed to contribute to the production of transient Frenkel pairs (t-FPs), because the empirical energy threshold to form HSSL in W is 230 eV per H *(32)*. Hence, only 1 % of the total H fluence was effectively responsible for HSSL formation *(32)*. The applied low H fluence was to avoid complex evolution of larger vacancy structures, ensuring that the free surface was the only available SIA sink throughout the entire plasma exposure. In other words, with the applied low H fluence the measured HSSL thickness should reflect the surface action range in absorbing SIAs during the plasma exposure. At higher H fluence, however, HSSL will propagate into larger depth beyond the intrinsic action range of the free surface. This is supposed to be realized via employing defects clusters in the shallow HSSL to absorb SIAs of t-FPs in larger depth (see **Fig. S1**). Under ions impact with 415 eV per H, the maximum kinetic energy transfer is about 9.0 eV in direct knock-on collisions ($H^+ \rightarrow W$). In the present work, the threshold energy of SIA formation in W (9.6 eV) (*46*) is taken as the lower limit of the threshold energy for t-FP production at the exposed surface ( $E_{d_0}^{t-FP}$ ). As the 9.0 eV kinetic energy transfer is even smaller than $E_{d_0}^{t-FP}$ , even t-FPs are not expected to form under these conditions. However, under the HiLaCoW effect the production of t-FPs becomes feasible at lower energies, as long as the collision occurs within the action range of the free surface in effectively absorbing SIAs. According to Ref. (*39*), the free surface can absorb all the SIAs within a depth of 2.5 ± 1 nm at 300 K, reflecting the initial thickness of HSSL without propagation into larger depth. At elevating temperature, the surface action range in absorbing SIA is supposed to be lower due the correspondingly lower surface energy (*45*). Since H thermally de-traps from the vacancies (see **Fig. S2**), the H occupancy level of vacancies drops with increasing temperature. Thus, to evaluate the actual number density of vacancies created by the plasma-exposures at 300-600 K, we treated the HSSL samples from 300-600 K with a second H plasma exposure at 300 K with 0 V bias and a large particle fluence of $1\times10^{22}$ $H^+$ $m^{-2}$ to compensate for any thermal loss of H occupancy. During the 2[nd] (zero-bias) plasma exposure, incident H ions carry only 15 eV of kinetic energy from the plasma potential, so no new vacancies are created but any existing H-unsaturated vacancy from the HSSL initially created at elevated temperatures will be (re)filled with H to the same occupancy level at 300 K (i.e., 5 atoms per vacancy (*41*)).

Hydrogen depth profiles in the hydrogen-supersaturated surface layers (HSSLs) formed on plasma-exposed tungsten were quantified by the resonant nuclear reaction $^{1}H$ ($^{15}N$, αγ) $^{12}C$ near the narrow 6.385 MeV energy resonance ($E_{res}$) (*42*). A $^{15}N^{2+}$ ion beam was incident normal to the sample surface in an ultra-high-vacuum chamber at the MALT tandem accelerator (University of Tokyo). The γ-ray yield from the nuclear reaction normalized to the number of incident $^{15}N$ ions ($Y$) is proportional to the H concentration ([H]) at a probing depth $d$, which is selected by the incident $^{15}$N ion energy ($E_i$) as $d = (E_i - E_{res})/S$, where $S$ is the stopping power ($S$) of the sample material for the ~6.4-MeV $^{15}$N ions. Quantification of hydrogen concentration (*42*) was achieved by calibrating the γ-detection efficiency using a Kapton ($(C_{22}H_{10}O_5N_2)_n$) foil as H-concentration standard (density: 1.45 g/cm$^3$, H-content: 2.284×10$^{22}$ cm$^{-3}$, and $S$ = 1.288 keV/nm), establishing the relation [H] (cm$^{-3}$) = $Y$ (cts/μC) × $S$ (keV/nm) / α, where α = 1.302×10$^{-19}$. For hydrogen-rich matrices, the stopping power increases according to Bragg's rule (*42*); for the present HSSL samples containing ~12 at.% H, the stopping power of pure W ($S$ = 4.010 keV/nm) increases by 1 % to $S$ = 4.051 keV/nm. Initial early-stage HSSL samples, as generated by the 1$^{st}$ plasma exposure with low-fluence, were installed from air without pre-measurement baking of the system and analyzed at room temperature under a vacuum < 1×10$^{-4}$ Pa. A dedicated chamber further enables in situ NRA during heating, allowing direct measurement of hydrogen outgassing up to 600 K, where a significant H degassing was observed (see **Fig. S2**). To ensure a consistent vacancy occupancy across samples towards a reliable analysis on the HSSL thickness, all HSSLs initially formed by the 1$^{st}$ plasma exposure at 300-600 K were subsequently exposed to a 2$^{nd}$ H plasma at 300 K for refilling.

## Supplementary Text

### Section 1 HSSL propagation upon higher $H^+$ fluence

Experimentally, the threshold energy transfer to create a HSSL in W was determined to be 5 ± 1 eV at 300 K (*32*). Collisions transferring kinetic energy above 5 eV to W lattice atoms shall be able to create t-FPs in the matrix. Highly mobile SIAs from these kinetically generated t-FPs are preferentially absorbed by nearby defect sinks resulting in local annihilation, whereas immobile vacancies survive and accumulate into extended defect structures that trap large quantities of H, forming the characteristic highly distorted HSSL (*31-33*). At low $H^+$ fluence (i.e., $<5\times10^{20}$ $m^{-2}$), as illustrated in **Fig. S1** and experimentally demonstrated in Ref. (*31*), the thickness of HSSL is limited within the action range (~2.5 ± 1 nm (*39*)) of the free surface as indicated by the dash-dotted line in **Fig. S1B**), because no other defects other than the free surface is acting as an intrinsic defect sink for SIA absorption and vacancy materialization. Upon increasing $H^+$ fluence, extended defects (i.e., vacancy clusters) can form at depth beyond the 2.5 ± 1 nm surface action range (*39*) and absorb newly generated SIAs from larger depths. Such kinetic vacancy relocation/diffusion accounts for the depth propagation of HSSL with increasing $H^+$ fluence. Upon an infinite H ion fluence at 415 eV, HSSL will eventually reach its maximum thickness that is determined by the threshold energy to create vacancies from t-FPs. For example, with 5 ± 1 eV as the threshold for kinetic t-FP generation in W at 300 K (*33*), the corresponding maximum HSSL thickness is ~14 nm according to SDTrimSP simulation (*48, 49*). Upon a finite $H^+$ fluence, however, the HSSL thickness, i.e., the effective depth in terms of t-FP generation, is solely determined by the kinetically-driven diffusion of vacancies at the applied exposure conditions, as explained below.

To track the propagation of HSSL as postulated above, we expose a set of high-purity poly-crystal W samples to H plasma with an ion energy of 415 eV at 300 K (see **Materials and**

**Methods**) to different H ion fluences. The H content (i.e., proportional to the vacancy number density) and thickness of the HSSLs are quantitatively evaluated with $^{1}$H-$^{15}$N nuclear reaction analysis (NRA) at a depth resolution of a few nanometers *(42)*. The NRA depth profiles of H in the HSSLs formed on the W surfaces after H plasma exposures to a fluence series are shown in **Fig. S1**). The NRA profile of a W sample before the plasma-exposure only shows absorbed H on the surface and no H content above the NRA sensitivity ($3 \times 10^{18}$ cm$^{-3}$) at depths beyond 5 nm (*32*), consistent with the low intrinsic defect density and low hydrogen solubility in pristine W materials. The thickness of the HSSL is estimated from the inflection point in the sigmoid fit curves to the measured H depth profiles (*32*), which is plotted as a function of the incident $H^+$ ion fluence in the inset figure in **Fig. S1B**). The square-root dependence of the HSSL thickness on the incident $H^+$ fluence in **Fig.S1B** (see the dashed line) clearly demonstrates the HSSL propagation via kinetically-driven vacancy diffusion. It is further postulated that the formation of vacancy clusters in the bulk acting as new SIA absorption sinks is dependent on sample temperature, which finally gives rise to different H concentration and layer thickness in the original HSSL formed initially at different temperatures.

**Section 2 Thermal stability of H trapping in HSSL**

To evaluate the thermal stability of H trapped at vacancy-type defects in the HSSL, the H occupancy level at the vacancies was directly measured as a function of sample temperature by $^{15}$N NRA on an HSSL/W sample formed at 300 K by a $3 \times 10^{24}$ m$^{-2}$ H ion exposure (ion energy 415 eV). The NRA incident $^{15}$N ion energy was fixed at ~6.40 MeV to probe H in the center of the HSSL in a depth of ~4 nm (cf. H profile in Fig. 4A for 300 K). **Fig. S2** shows the NRA γ-yield from the residual H content in the HSSL after the sample was heated in vacuum ($2 \times 10^{-5}$ Pa) for 1 minute at each of the plotted temperatures in the range from 300 to 600 K. As evident from the figure, the H content gradually decreases in a broad decay due to thermal de-trapping of H from the vacancy defects, showing only a slight decrease between 300-400 K followed by a more rapid decay between 400-600 K. This temperature dependence demonstrates the fact that the H contents in the as-prepared HSSL samples at various temperatures are different. The observed temperature dependence of remaining H is consistent with TDS spectra from HSSL samples, which show a small $H_2$ desorption peak at 460 K and a main desorption feature centered around 540 K (*50*). With this knowledge, the refilling plasma exposure at 300 K was performed for the HSSL samples initially formed at temperatures between 300-600 K and gave rise to the same H occupancy in all the HSSL samples.

**Fig. S1.**

***The propagation of hydrogen supersaturated surface layer (HSSL) in W exposed to H plasma with different $H^+$ fluence at 415 eV ion energy.*** *A) H depth profiles in the fluence series HSSL samples, as measured by $^{1}H$-$^{15}N$ nuclear reaction analysis (NRA). The upper depth scale is formed by corresponding the 0 nm to the $^{1}H$-$^{15}N$ resonance at 6.385 MeV in the bottom $^{15}N$ ion energy scale and further calculated according to the stopping power (31, 42). Note that experimental uncertainty applies for all the curves, but for the sake of general clarity, only one curve is inserted with experimental error bar. B) The measured HSSL thickness as a function of the incident 415 eV $H^+$ fluence up to $1 \times 10^{22}$ $m^{-2}$. Two data points from higher $H^+$ fluence are adopted from the present work and also Ref. (31). The dash-dotted line indicates the intrinsic action range of the free surface in absorbing SIA from the depth, which was measured to be 2.5 ± 1 nm for W at 300 K in Ref. (39). The identified square root dependence (as denoted by the dashed line) of HSSL thickness with $H^+$ fluence demonstrates the kinetically-driven diffusion of vacancies that facilitate HSSL propagation into larger depth.*

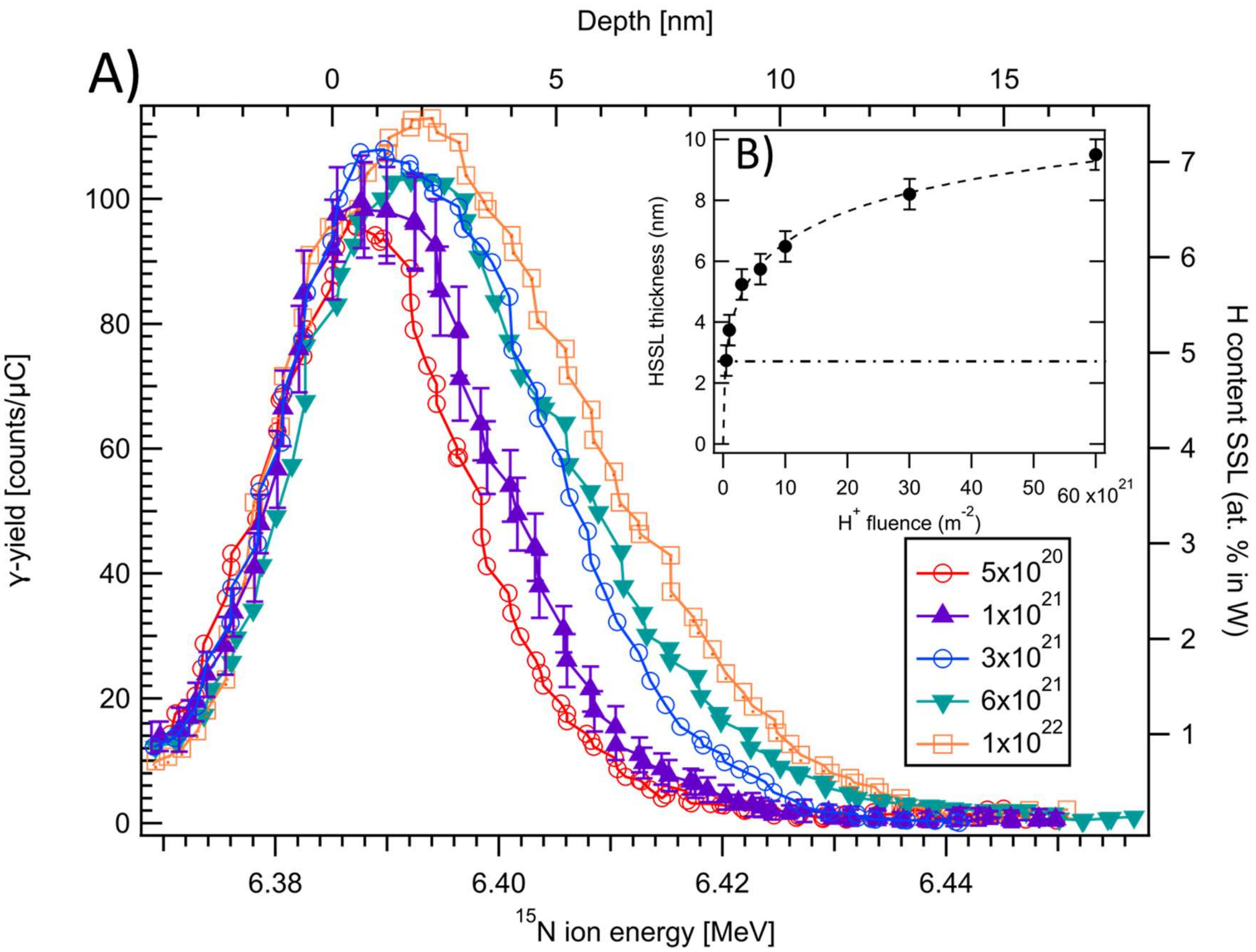

**Fig. S2.**

***Thermal stability measurement the hydrogen supersaturated surface layer (HSSL).*** *Temperature-dependent NRA γ-yield from hydrogen probed in the center of the HSSL (in 4 nm depth) on a tungsten sample prepared (by exposure to 3 × $10^{24}$ $m^{-2}$ H ions with 415 eV) at 300 K, cf. Fig. 4A. Each NRA measurement was performed after heating the sample in vacuum (2 × $10^{-5}$ Pa) for 1 minute at the plotted temperatures. A significant decrease of the NRA γ-yield has been detected at elevated temperatures.*

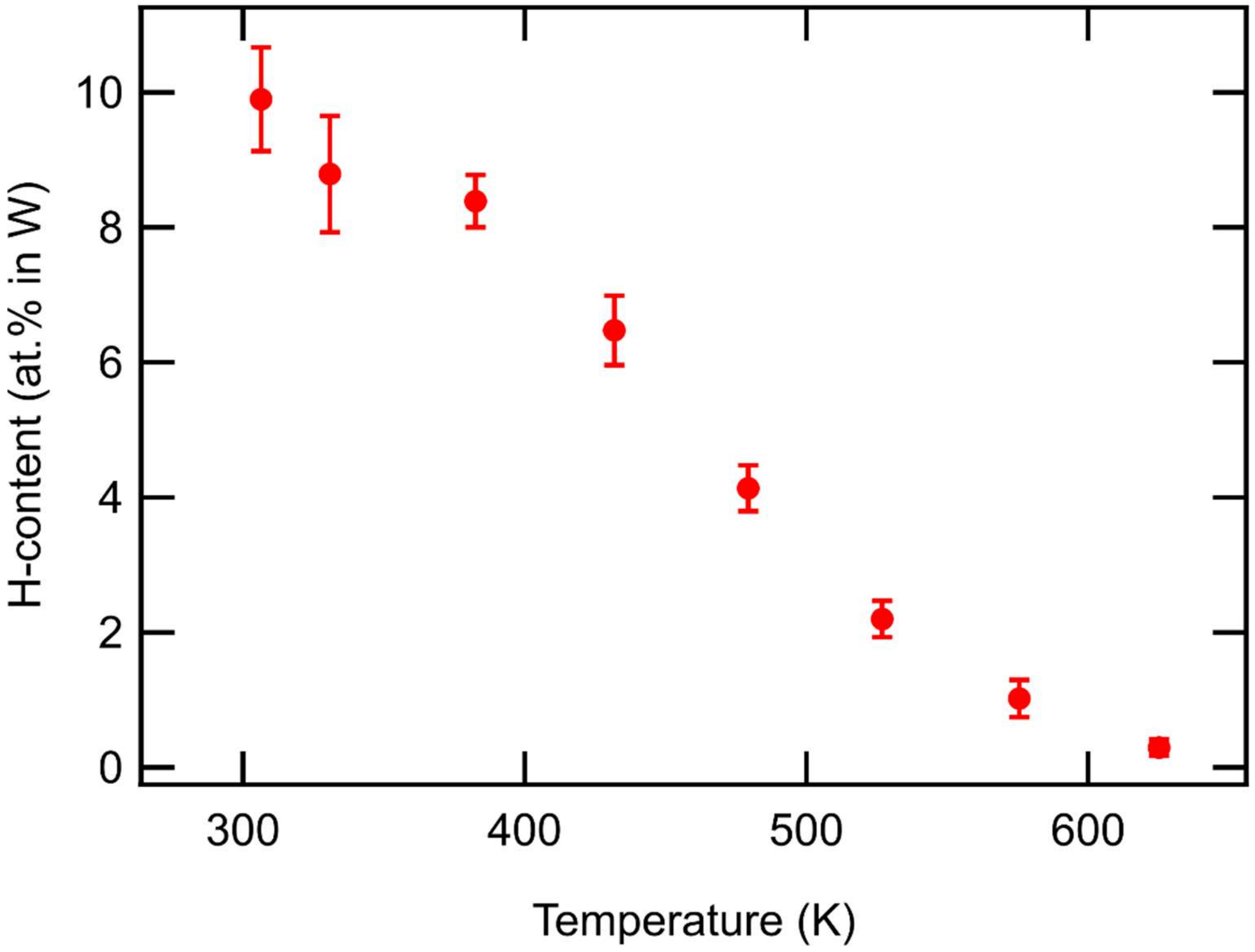

**References (47-50)**